# Multiresolution finite element method based on a new locking-free rectangular Mindlin plate element

YiMing Xia

(Civil Engineering Department, Nanjing University of Aeronautics and Astronautics, Nanjing 210016, China)

(Email: xym4603@sina.com)

*Abstract*—

A locking-free rectangular Mindlin plate element with a new multi-resolution analysis (MRA) is proposed and a new finite element method is hence presented. The MRA framework is formulated out of a mutually nesting displacement subspace sequence whose basis functions are constructed of scaling and shifting on the element domain of basic node shape function. The basic node shape function is constructed by extending the node shape function of a traditional Mindlin plate element to other three quadrants around the coordinate zero point. As a result, a new rational MRA concept together with the resolution level (RL) is constituted for the element. The traditional 4-node rectangular Mindlin plate element and method is a mono-resolution one and also a special case of the proposed element and method. The meshing for the monoresolution plate element model is based on the empiricism while the RL adjusting for the multiresolution is laid on the rigorous mathematical basis. The analysis clarity of a plate structure is actually determined by the RL, not by the mesh. Thus, the accuracy of a plate structural analysis is replaced by the clarity, the irrational MRA by the rational and the mesh model by the RL that is the discretized model by the integrated.

## 1. Introduction

Multi-resolution analysis (MRA) is a popular technique that has been applied in many domains such as the signal and image processing, the damage detection and health monitoring, the differential equation solution, etc. However, in the field of computational mechanics, the MRA has not been, in a real sense, fully utilized in the numerical solution of engineering problems either by the traditional finite element method (FEM)[1] or by other methods such as the wavelet finite element method (WFEM)[2, 3], the meshfree method (MFM)[4, 5] and the natural element method (NEM)[6, 7] etc.

As is commonly known of the FEM, owing to the invariance of node number a single finite element contains, the finite element can be regarded as a monoresolution one from a MRA point of view and the FEM structural analysis is usually not associated with the MRA concept. The MRA seems to be rarely used when the FEM is employed to numerical analysis. However, it is, in fact, by means of meshing and re-meshing in which a cluster of monoresolution finite elements are assembled together artificially that the rough MRA is executed by the FEM. As we can see, in overall analysis process of a structure by the FEM, there is no mathematical foundation for the traditional finite element meshing and the finite elements are assembled together artificially. The traditional finite element model has to be re-meshed until sufficient accuracy is reached, which leads to the low computation efficiency or convergent rate. The deficiency of the FEM becomes much explicit in the accurate computation of structural problems with local steep gradient such as material nonlinear [8, 9], local damage and crack [10, [11], impacting and exploding problems [12, 13].

The great efforts have been made over the past thirty years to overcome the drawbacks of the FEM with many improved methods to come up, such as WFEM, MFM and NEM etc, which open up a transition from the monoresolution finite element method to the multiresolution finite element method featured with adjustable element node number. Although these MRA methods have illustrated their powerful capability and computational efficiency in dealing with some problems, they always have such major inherent deficiencies as the complexity of shape function construction, the absence of the Kronecker delta property of the shape function and the lack of a rigorous mathematical basis for the MRA, which make the treatment of element boundary condition complicated and the selection of element node layout empirical, that substantially reduce computational efficiency. Hence, these MRA methods have never found a wide application in engineering practice just as the FEM. In fact, they can be viewed as the intermediate products in the transition of the FEM from the monoresolution to the multiresolution.

The deficiencies of all those MRA methods can be eliminated by the introduction of a new multiresolution finite element method in this paper. With respect to Mindlin plate element in the finite element stock, a new multiresolution locking-free rectangular Mindlin plate element is formulated by the MRA based on a displacement subspace sequence which is constituted by translated and scaled version as subspace basis functions of the basic node shape functions for a locking-free rectangular Mindlin plate element. The basic node shape function is then constructed from shifting to other three quadrants around a specific node of a basic element in one quadrant and joining the corresponding node shape functions of four elements at the specific node. Hence, the node shape function construction is quite simple and clear. In addition, the proposed element method possesses a simple, clear and rigorous mathematical basis for MRA, which endows the proposed element with the resolution level (RL) that can be modulated to freely change the element node number and position in the element, adjusting structural analysis accuracy accordingly. As a result, the proposed element method can bring about substantial improvement of the computational efficiency in the structural analysis when compared with the corresponding FEM or other MRA methods.

## 2. Basic node shape functions

A rectangular Mindlin plate element considered in this paper is shown as Fig1. The following is the crucial step of constructing basic node shape functions in formulation of the multi-resolution locking-free rectangular Mindlin plate element.

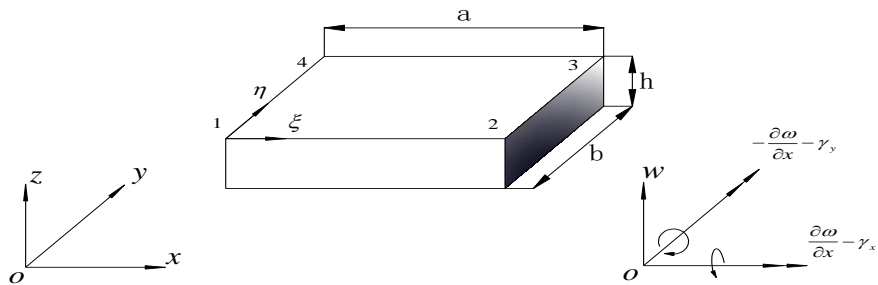

**Fig 1.** A rectangular Mindlin plate element

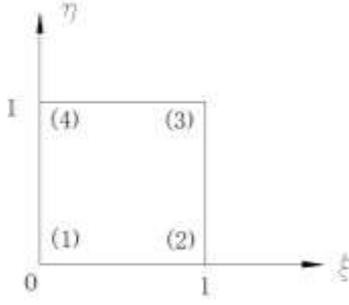 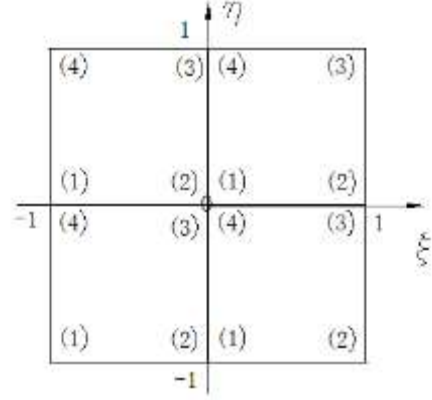

**Fig 2.** The node shape function domain of Mindlin plate element

**Fig 3.** The extended shape function domain for the basic node at the point coordinate of (0, 0)

The displacement of a classical rectangular Mindlin plate element shown in Fig.1 can be easily acquired and concisely expressed in terms of natural coordinates as follows[14]:

$$\begin{cases} w = \sum_{i=1}^{4}\left[ N_i^b w_i + b N_{xi}^b (\theta_{xi} + \gamma_{xi}) + a N_{yi}^b (\theta_{yi} + \gamma_{yi}) \right] \\ \gamma_x = \sum_{i=1}^{4} N_i^0 \gamma_{xi} \\ \gamma_y = \sum_{i=1}^{4} N_i^0 \gamma_{yi} \end{cases} \quad (1)$$

where $w$, $\gamma_x, \gamma_y$ are the transverse displacement and the shear angles around the $x,y$ axis directions at an arbitrary point of the element respectively. $w_i, \theta_{xi}, \theta_{yi}, \gamma_{xi}, \gamma_{yi}$ are the transverse, the rotational displacements and the shear angles at node $i$ of the element respectively. $N_i^0$, $N_i^b$, $N_{xi}^b$, $N_{yi}^b$ are the conventional shape functions at the node $i$.

The conventional shape functions $N_i^0, N_i^b, N_{xi}^b, N_{yi}^b$ are defined on the domain of $[0,1]^2$ as follows

$$\begin{cases} N_1^0 = (1-\xi)(1-\eta) \\ N_2^0 = \xi(1-\eta) \\ N_3^0 = \xi\eta \\ N_4^0 = (1-\xi)\eta \end{cases} \quad (2)$$

$$\begin{cases} N_1^b(\xi,\eta) = (1-\xi)(1-\eta)(\xi+\eta-2\xi^2-2\eta^2+1) \\ N_2^b(\xi,\eta) = \xi(1-\eta)(3\xi+\eta-2\xi^2-2\eta^2) \\ N_3^b(\xi,\eta) = \xi\eta(3\xi+3\eta-2\xi^2-2\eta^2-1) \\ N_4^b(\xi,\eta) = \eta(1-\xi)(\xi+3\eta-2\xi^2-2\eta^2) \end{cases} \quad (3)$$

$$\begin{cases} N^b_{x1}(\xi,\eta) = \eta(1-\eta)^2(1-\xi) \\ N^b_{x2}(\xi,\eta) = \eta(1-\eta)^2 \xi \\ N^b_{x3}(\xi,\eta) = -\eta^2(1-\eta)\xi \\ N^b_{x4}(\xi,\eta) = -\eta^2(1-\eta)(1-\xi) \end{cases} \quad (4)$$

$$\begin{cases} N^b_{y1}(\xi,\eta) = \xi(1-\xi)^2(1-\eta) \\ N^b_{y2}(\xi,\eta) = -\xi^2(1-\xi)(1-\eta) \\ N^b_{y3}(\xi,\eta) = -\xi^2(1-\xi)\eta \\ N^b_{y4}(\xi,\eta) = \xi(1-\xi)^2 \eta \end{cases} \quad (5)$$

in which $\xi = \dfrac{x}{a}, \eta = \dfrac{y}{b}$

Meanwhile, the shear angle $\gamma_{xi}, \gamma_{yi}$ at node $i$ can be read respectively as [14]:

$$\begin{cases} \gamma_{x1} = (\delta_2/b)[2(w_1-w_4)-b(\theta_{x1}+\theta_{x4})] \\ \gamma_{x2} = (\delta_2/b)[2(w_2-w_3)-b(\theta_{x2}+\theta_{x3})] \\ \gamma_{x3} = (\delta_2/b)[2(w_2-w_3)-b(\theta_{x2}+\theta_{x3})] \\ \gamma_{x4} = (\delta_2/b)[2(w_1-w_4)-b(\theta_{x1}+\theta_{x4})] \end{cases} \quad (6)$$

$$\begin{cases} \gamma_{y1} = (\delta_1/a)[2(w_2-w_1)-a(\theta_{y1}+\theta_{y2})] \\ \gamma_{y2} = (\delta_1/a)[2(w_2-w_1)-a(\theta_{y1}+\theta_{y2})] \\ \gamma_{y3} = (\delta_1/a)[2(w_3-w_4)-a(\theta_{y3}+\theta_{y4})] \\ \gamma_{y4} = (\delta_1/a)[2(w_3-w_4)-a(\theta_{y3}+\theta_{y4})] \end{cases} \quad (7)$$

in which the parameters $\delta_1 = \dfrac{(h/a)^2}{5/6(1-\mu)+2(h/a)^2}$  $\delta_2 = \dfrac{(h/b)^2}{5/6(1-\mu)+2(h/b)^2}$ , $h$ is denoted as the thickness of the plate, $\mu$ as the Poisson's ratio.

Hence, substituting (6), (7) into (1), the following equations can be obtained as:

$$\begin{cases} w = w_b + w_s \\ w_b = \sum_{i=1}^{4} \left[ N_i^b w_i + b N_{xi}^b \theta_{xi} + a N_{yi}^b \theta_{yi} \right] \\ w_s = \sum_{i=1}^{4} \left[ N_i^s w_i + b N_{xi}^s \theta_{xi} + a N_{yi}^s \theta_{yi} \right] \\ \gamma_x = \sum_{i=1}^{4} \left[ N_i^{\gamma_x} w_i + b N_{xi}^{\gamma_x} \theta_{xi} \right] \\ \gamma_y = \sum_{i=1}^{4} \left[ N_i^{\gamma_y} w_i + a N_{yi}^{\gamma_y} \theta_{yi} \right] \end{cases} \quad (8)$$

where $w_b, w_s$ are the transverse displacement fields caused by the bending and the shear deformation respectively.

In addition, there exists the following relationship

$$\theta_x = \frac{\partial w}{\partial y} - \gamma_x, \quad \theta_y = -\frac{\partial w}{\partial x} - \gamma_y \quad (9)$$

As to the proposed element, the various shape functions, regarding to a node at the point coordinate of (0, 0) as shown in the Fig 2. initially defined on the domain of $[0, 1]^2$ or viewed as for a 1/4 split node, should be extended to the domain of $[-1,1]^2$ for a full node by means of shifting the element around the node (1) vertically, horizontally and obliquely respectively to the other three quadrants, thus covering the eight nodes adjacent to the basic node as displayed in Fig 3, finally the basic shape functions for the full node at the point coordinate of (0, 0) can be defined as follows:

$$\phi_1^I(\xi,\eta) := \begin{cases} N_1^I(\xi,\eta) & \xi \in [0,1], \eta \in [0,1] \\ N_2^I(1+\xi,\eta) & \xi \in [-1,0], \eta \in [0,1] \\ N_3^I(1+\xi,1+\eta) & \xi \in [-1,0], \eta \in [-1,0] \\ N_4^I(\xi,1+\eta) & \xi \in [0,1], \eta \in [-1,0] \\ 0 & \xi \notin [-1,1], \eta \notin [-1,1] \end{cases} \quad (10)$$

$$\phi_2^I(\xi,\eta) := \begin{cases} N_{x1}^I(\xi,\eta) & \xi \in [0,1], \eta \in [0,1] \\ N_{x2}^I(1+\xi,\eta) & \xi \in [-1,0], \eta \in [0,1] \\ N_{x3}^I(1+\xi,1+\eta) & \xi \in [-1,0], \eta \in [-1,0] \\ N_{x4}^I(\xi,1+\eta) & \xi \in [0,1], \eta \in [-1,0] \\ 0 & \xi \notin [-1,1], \eta \notin [-1,1] \end{cases} \quad (11)$$

$$\phi_3^I(\xi,\eta) := \begin{cases} N_{y1}^I(\xi,\eta) & \xi \in [0,1], \eta \in [0,1] \\ N_{y2}^I(1+\xi,\eta) & \xi \in [-1,0], \eta \in [0,1] \\ N_{y3}^I(1+\xi,1+\eta) & \xi \in [-1,0], \eta \in [-1,0] \\ N_{y4}^I(\xi,1+\eta) & \xi \in [0,1], \eta \in [0,-1] \\ 0 & \xi \notin [-1,1], \eta \notin [-1,1] \end{cases} \quad (12)$$

where the superscript $I$ be denoted as $b$, $s$, $\gamma_x$ and $\gamma_y$ respectively.

The Kronecker delta property holds for the basic node shape functions $\phi_1^b(\xi,\eta)$, $\phi_2^b(\xi,\eta)$, $\phi_3^b(\xi,\eta)$:

$$\begin{cases} \phi_1^b(0,0)=1 & \phi_1^b(8nodes)=0 & \dfrac{\partial \phi_1^b(0,0)}{\partial \xi}=0 \\ \dfrac{\partial \phi_1^b(0,0)}{\partial \eta}=0 & \dfrac{\partial \phi_1^b(8nodes)}{\partial \xi}=0 & \dfrac{\partial \phi_1^b(8nodes)}{\partial \eta}=0 \\ \phi_2^b(0,0)=0 & \phi_2^b(8nodes)=0 & \dfrac{\partial \phi_2^b(0,0)}{\partial \xi}=0 \\ \dfrac{\partial \phi_2^b(0,0)}{\partial \eta}=1 & \dfrac{\partial \phi_2^b(8nodes)}{\partial \eta}=0 & \dfrac{\partial \phi_2^b(8nodes)}{\partial \xi}=0 \\ \phi_3^b(0,0)=0 & \phi_3^b(8nodes)=0 & \dfrac{\partial \phi_3^b(0,0)}{\partial \xi}=-1 \\ \dfrac{\partial \phi_3^b(0,0)}{\partial \eta}=0 & \dfrac{\partial \phi_3^b(8nodes)}{\partial \xi}=0 & \dfrac{\partial \phi_3^b(8nodes)}{\partial \eta}=0 \end{cases} \quad (13)$$

The basic full node shape functions $\phi_1^b(\xi,\eta)$, $\phi_2^b(\xi,\eta)$ and $\phi_3^b(\xi,\eta)$ are shown in Figs. 4.*a*, *b*, *c* respectively.

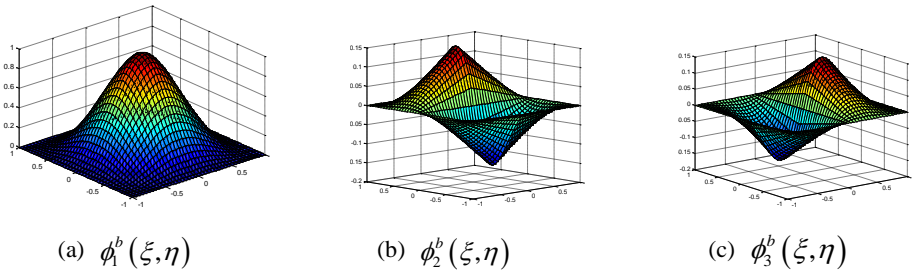

(a) $\phi_1^b(\xi,\eta)$      (b) $\phi_2^b(\xi,\eta)$      (c) $\phi_3^b(\xi,\eta)$

**Fig 4.** The basic node shape functions $\phi_1^b(\xi,\eta)$, $\phi_2^b(\xi,\eta)$, $\phi_3^b(\xi,\eta)$ on the domain of $[-1,1] \times [-1,1]$

## 3. Displacement subspace sequence

In order to carry out a MRA of a rectangular Mindlin plate structure, the mutual nesting displacement subspace sequence for a rectangular Mindlin plate element should be established. In this paper, a totally new technique is proposed to construct the MRA which is based on the concept that a subspace sequence (multi-resolution subspaces) can be formulated by subspace basis function vectors at different resolution levels whose elements-scaling function vector can be constructed by scaling and shifting on the domain $[0,1]^2$ of the basic full node shape functions. As a result, the displacement subspace basis function vector at an arbitrary resolution level (RL) of $(m+1)\times(n+1)$ for a rectangular Mindlin plate element with the domain of $a\times b$ is formulated as follows:

$$\Pi_{mn} = \begin{Bmatrix} \Psi_{mn} \\ \Omega_{mn} \end{Bmatrix} \quad \begin{cases} \Psi_{mn} = \begin{bmatrix} \Phi_{mn,00} & \cdots & \Phi_{mn,rs} & \cdots & \Phi_{mn,mn} \end{bmatrix} \\ \Omega_{mn} = \begin{bmatrix} \omega_{mn,00} \cdots & \omega_{mn,rs} \cdots & \omega_{mn,mn} \end{bmatrix} \end{cases} \tag{14}$$

where $\Phi_{mn,rs} = \begin{bmatrix} \phi^b_{1mn,rs} + \phi^s_{1mn,rs} & l_{ey}(\phi^b_{2mn,rs} + \phi^s_{2mn,rs}) & l_{ex}(\phi^b_{3mn,rs} + \phi^s_{3mn,rs}) \end{bmatrix}$,

$\omega_{mn,rs} = \begin{bmatrix} \omega^{\gamma_x}_{mn,rs} \\ \omega^{\gamma_y}_{mn,rs} \end{bmatrix} = \begin{bmatrix} \phi^{\gamma_x}_{1mn,rs} & l_{ey}\phi^{\gamma_x}_{2mn,rs} & 0 \\ \phi^{\gamma_y}_{1mn,rs} & 0 & l_{ex}\phi^{\gamma_y}_{3mn,rs} \end{bmatrix}$ are the scaling basis function vectors,

$\phi^I_{1mn,rs} = \phi^I_1(m\xi - r, n\eta - s)$, $\phi^I_{2mn,rs} = \phi^I_2(m\xi - r, n\eta - s)$, $\phi^I_{3mn,rs} = \phi^I_3(m\xi - r, n\eta - s)$, $m$, $n$ denoted as the positive integers, that are the scaling parameters in $\xi, \eta$ directions respectively. $r$, $s$ as the positive integers, the node position parameters, that is $r = 0,1,2,3\cdots m$, $s = 0,1,2,3\cdots n$. $l_{ex} = \dfrac{a}{m}$, $l_{ey} = \dfrac{b}{n}$,

$$\delta_{1m} = \frac{(mh/a)^2}{5/6(1-\mu)+2(mh/a)^2} \quad \delta_{2n} = \frac{(nh/b)^2}{5/6(1-\mu)+2(nh/b)^2}$$

Here, $(m\xi - r) \in [-1,1], (n\eta - s) \in [-1,1], \xi \in [0,1], \eta \in [0,1]$.

It is seen from Eq. (14) that the nodes for the scaling process are equally spaced on the element domain of $[0,1]^2$ with a step size of $1/m$ in $\xi$ and $1/n$ in $\eta$ directions respectively.

Scaling of the basic node shape functions on the domain of $[-1, 1]^2$ (precisely on the domain of $\left[-\dfrac{1}{m},\dfrac{1}{m}\right] \times \left[-\dfrac{1}{n},\dfrac{1}{n}\right]$) and then shifting to other nodes $\left(\dfrac{r}{m},\dfrac{s}{n}\right)$ on the element domain of $[0,1]^2$ will produce the various node shape functions that are shown in the Fig 5. at the RL of $2\times 2$, $3\times 3$.

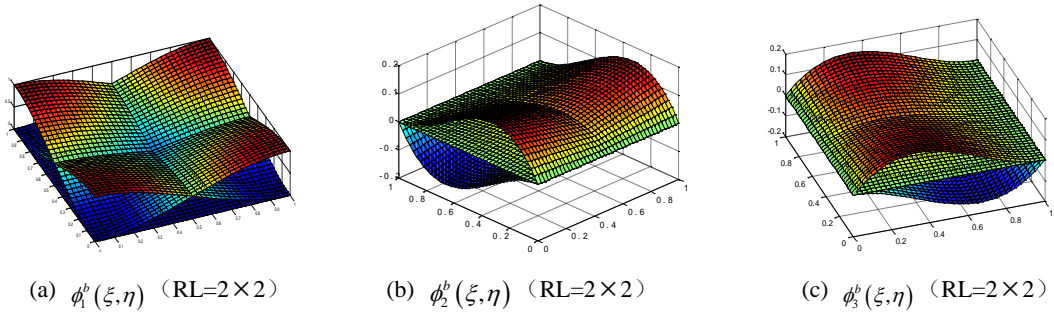

(a) $\phi^b_1(\xi,\eta)$ (RL=2×2)     (b) $\phi^b_2(\xi,\eta)$ (RL=2×2)     (c) $\phi^b_3(\xi,\eta)$ (RL=2×2)

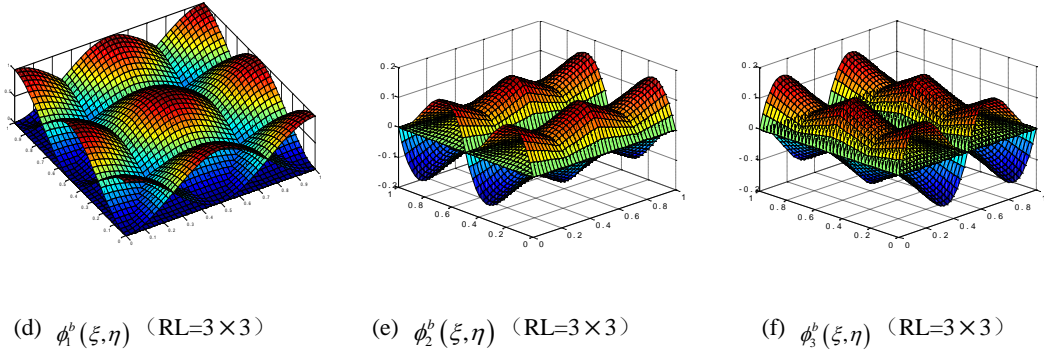

(d) $\phi_1^b(\xi,\eta)$ （RL=3×3）    (e) $\phi_2^b(\xi,\eta)$ （RL=3×3）    (f) $\phi_3^b(\xi,\eta)$ （RL=3×3）

**Fig5**. The scaled and shifted version of the basic node shape functions $\phi_1^b(\xi,\eta), \phi_2^b(\xi,\eta), \phi_3^b(\xi,\eta)$ on the element domain of $[0,1]\times[0,1]$

Since the elements in the basis functions are linearly independent with the various scaling and the different shifting parameters, the subspaces in the subspace sequence can be established and are mutually nested, thus formulating a MRA framework, that is

$$\begin{cases} \mathbf{W}_{mn} = \begin{bmatrix} V_{11} \cdots & V_{ij} \cdots & V_{mn} \end{bmatrix} \\ V_{ij} := span\{\mathbf{\Pi}_{ij} : i, j \in Z\} \\ V_{ij} \subset V_{i(j+1)} \quad V_{ij} \subset V_{(i+1)j} \quad V_{ij} \subset V_{(i+1)(j+1)} \end{cases} \quad (15)$$

where $Z$ denoted as the positive integers, $V_{ij}$ as displacement subspace at the resolution level of $(i+1)\times(j+1)$.

Thus, it can be seen that the mutually nesting displacement subspace sequence $\mathbf{W}_{mn}$ can be taken as a simple, clear and solid mathematical foundation for the MRA framework.

Based on the MRA, the deflection of a rectangular Mindlin plate element in the displacement subspace at the RL of $(m+1)\times(n+1)$ can be defined as follows

$$\mathbf{d}_{mn} = \begin{Bmatrix} w_{mn} \\ \gamma_{xmn} \\ \gamma_{ymn} \end{Bmatrix} = \begin{Bmatrix} \mathbf{\Psi}_{mn} \\ \mathbf{\Omega}_{mn} \end{Bmatrix} \mathbf{a}_{mn}^e \quad (16)$$

where $\mathbf{a}_{mn}^e = \begin{bmatrix} [w_{00},\theta_{x00},\theta_{y00}] \cdots & [w_{rs},\theta_{xrs},\theta_{yrs}] \cdots & [w_{mn},\theta_{xmn},\theta_{ymn}] \end{bmatrix}^T$, $w_{rs},\theta_{xrs},\theta_{yrs}$ are the transverse, rotational displacements respectively at the element node $\left(\dfrac{r}{m},\dfrac{s}{n}\right)$.

It can be seen that the proposed multi-resolution element is a meshfree one whose nodes are uniformly scattered at each coordinate respectively, node number and position fully determined by the *RL*. When the *RL*=2×2, that is a traditional 4-node rectangular Mindlin plate element, Eq. (16) will be reduced to Eq. (1). Hence, the traditional 4-node rectangular Mindlin plate element can be regarded as a mono-resolution one and also a special case of the multi-resolution rectangular Mindlin plate element.

## 4. Multiresolution rectangular Mindlin plate element formulation

According to the classical assumption of a Mindlin plate theory, the generalized function of potential energy in a displacement subspace at the resolution level of $(m+1)\times(n+1)$ for a rectangular Mindlin plate element with the domain of $a\times b$ can be listed as:

$$\Pi_p(V_{mn}) = \frac{1}{2}\int_0^a\int_0^b \boldsymbol{\kappa}_{mn}^T \mathbf{D}_b \boldsymbol{\kappa}_{mn} dxdy - \frac{1}{2}\int_0^a\int_0^b \boldsymbol{\gamma}_{mn}^T \mathbf{D}_s \boldsymbol{\gamma}_{mn} dxdy \\ - \int_0^a\int_0^b q w_{mn} dxdy - \sum_i Q_i w_{mni} \tag{17}$$

where $\boldsymbol{\kappa}_{mn} = -\begin{bmatrix} \dfrac{\partial^2 w_{mn}}{\partial x^2} - \dfrac{\partial \gamma_{xmn}}{\partial x} \\ \dfrac{\partial^2 w_{mn}}{\partial y^2} - \dfrac{\partial \gamma_{ymn}}{\partial y} \\ 2\dfrac{\partial^2 w_{mn}}{\partial x\partial y} - \left(\dfrac{\partial \gamma_{ymn}}{\partial y} + \dfrac{\partial \gamma_{xmn}}{\partial x}\right) \end{bmatrix}$, $\boldsymbol{\gamma}_{mn} = \begin{bmatrix} \gamma_{xmn} \\ \gamma_{ymn} \end{bmatrix}$, $\mathbf{D}_b = C_b\begin{bmatrix} 1 & \mu & 0 \\ \mu & 1 & 0 \\ 0 & 0 & (1-\mu)/2 \end{bmatrix}$,

$\mathbf{D}_s = C_s\begin{bmatrix} (1-\mu)/2 & 0 \\ 0 & (1-\mu)/2 \end{bmatrix}$, $C_b = Eh^3/12(1-\mu^2)$,

$C_s = kGh = 5Eh/12(1+\mu)$. $E$ is the material Young modulus, $h$ the thickness of the element, $\mu$ the Poisson's ratio, $q$ distributed transverse loadings, $Q_i$ the lump transverse loadings.

$$\begin{cases} \boldsymbol{\kappa}_{mn} = \left[\mathbf{B}_{00}^b, \cdots \mathbf{B}_{rs}^b, \cdots \mathbf{B}_{mn}^b\right]\mathbf{a}_{mn}^e \\ \boldsymbol{\gamma}_{mn} = \left[\mathbf{B}_{00}^s, \cdots \mathbf{B}_{rs}^s, \cdots \mathbf{B}_{mn}^s\right]\mathbf{a}_{mn}^e \end{cases} \tag{18}$$

式中

$$\mathbf{B}_{rs}^b = -\left[\dfrac{\partial^2 \Phi_{mn,rs}}{\partial x^2} - \dfrac{\partial \omega_{mn,rs}^{\gamma_x}}{\partial x} \quad \dfrac{\partial^2 \Phi_{mn,rs}}{\partial y^2} - \dfrac{\partial \omega_{mn,rs}^{\gamma_y}}{\partial y} \quad 2\dfrac{\partial^2 \Phi_{mn,rs}}{\partial x\partial y} - \left(\dfrac{\partial \omega_{mn,rs}^{\gamma_x}}{\partial y} + \dfrac{\partial \omega_{mn,rs}^{\gamma_y}}{\partial x}\right)\right]^T,$$

$$\mathbf{B}_{rs}^s = \left[\omega_{mn,rs}^{\gamma_x} \quad \omega_{mn,rs}^{\gamma_y}\right]^T.$$

Substitute Eq.(15) into Eq.(16), the concise expression can be obtained after reassembling as follows:

$$\Pi_p(V_{mn}) = \frac{1}{2}\mathbf{a}_{mn}^{eT}\mathbf{K}_{mn}^e\mathbf{a}_{mn}^e - \mathbf{a}_{mn}^{eT}\mathbf{f}_{mn}^e - \mathbf{a}_{mn}^{eT}\mathbf{F}_{mn}^e \tag{19}$$

where $\mathbf{K}_{mn}^e$ is the element stiffness matrix, $\mathbf{f}_{mn}^e$ the element distributed loading column vector, $\mathbf{F}_{mn}^e$ the element lump loading column vector.

According to the potential energy minimization principle, let $\delta\Pi_p(V_{mn}) = 0$, the shell element equilibrium equations can be obtained as follows,

$$\mathbf{K}_{mn}^{e}\mathbf{a}_{mn}^{e} = \mathbf{f}_{mn}^{e} + \mathbf{F}_{mn}^{e} \qquad (20)$$

The element expressions of the stiffness matrix $\mathbf{K}_{mn}^{e}$, and the loading column vectors $\mathbf{f}_{mn}^{e}$, $\mathbf{F}_{mn}^{e}$ can be given as follows:

$$\mathbf{K}_{mn}^{e} = \begin{bmatrix} \mathbf{k}_{00}^{00} & \cdots & \mathbf{k}_{rs}^{00} & \cdots & \mathbf{k}_{mn}^{00} \\ \vdots & & \vdots & & \vdots \\ \mathbf{k}_{00}^{rs} & \cdots & \mathbf{k}_{rs}^{rs} & \cdots & \mathbf{k}_{ij}^{rs} \\ \vdots & & \vdots & & \vdots \\ \mathbf{k}_{00}^{mn} & \cdot & \mathbf{k}_{rs}^{mn} & \cdot & \mathbf{k}_{mn}^{mn} \end{bmatrix} \qquad (21)$$

where the superscript denoted as the row number of the matrix and the subscript as the aligned element node numbering ($r$, $s$). In terms of the properties of the extended shape functions, we have

$$\begin{cases} \mathbf{k}_{rs}^{rs} = \sum_{\substack{|c-r|\leq 1 \\ |d-s|\leq 1}} \mathbf{k}_{cd,rs} \\ \mathbf{k}_{rs}^{rs} = \mathbf{k}_{cd,rs} = 0, when\ |c-r|>1, |d-s|>1 \end{cases} \qquad (22)$$

in which $\mathbf{k}_{cd,rs}$ is the coupled node stiffness matrix relating the node ($c$, $d$) to ($r$, $s$).

$$\mathbf{k}_{cd,rs} = ab\int_{0}^{1}\int_{0}^{1} \begin{bmatrix} \mathbf{B}_{cd}^{b} & \mathbf{B}_{cd}^{s} \end{bmatrix}^{T} \begin{bmatrix} \mathbf{D}_{b} & 0 \\ 0 & \mathbf{D}_{s} \end{bmatrix} \begin{bmatrix} \mathbf{B}_{rs}^{b} & \mathbf{B}_{rs}^{s} \end{bmatrix} d\xi d\eta \qquad (23)$$

$$\begin{cases} \mathbf{f}_{mn,rs}^{er} = a.b\int_{0}^{1}\int_{0}^{1} \begin{bmatrix} \mathbf{\Phi}_{mn,rs} \end{bmatrix}^{T} q d\xi d\eta \\ \mathbf{F}_{mn,rs}^{er} = \sum_{i} \begin{bmatrix} \mathbf{\Phi}_{mn,rs}(m\xi_{i}-r, n\eta_{i}-s) \end{bmatrix}^{T} Q_{i} \end{cases} \qquad (24)$$

where $\xi_{i}$, $\eta_{i}$ is the local coordinate at the locations the lump loading acting on.

## 5. Transformation matrix

In order to carry out structural analysis, the element stiffness and mass matrices $\mathbf{K}_{mn}^{e}$ the loading column vectors $\mathbf{f}_{mn}^{e}$, $\mathbf{F}_{mn}^{e}$ should be transformed from the element local coordinate system ($xyz$) to the structural global coordinate system ($XYZ$). The transforming relations from the local to the global are defined as follows:

$$\mathbf{K}_{mn}^{i} = \mathbf{T}_{mn}^{eT}\mathbf{K}_{mn}^{e}\mathbf{T}_{mn}^{e} \qquad (25)$$

$$\mathbf{f}_{mn}^{i} = \mathbf{T}_{mnl}^{eT}\mathbf{f}_{mn}^{e} \tag{26}$$

$$\mathbf{F}_{mn}^{i} = \mathbf{T}_{mnl}^{eT}\mathbf{F}_{mn}^{e} \tag{27}$$

where $\mathbf{K}_{mn}^{i}$ is the element stiffness matrix, $\mathbf{f}_{mn}^{i}, \mathbf{F}_{mn}^{i}$ the element loading column vectors under the global coordinate system. $\mathbf{T}_{mn}^{e}$ is the element transformation matrix defined as follows:

$$\mathbf{T}_{mn}^{e} = \begin{bmatrix} \lambda_{11} & & & & \mathbf{0} \\ & \ldots & & & \\ & & \lambda_{ij} & & \\ & & & \ldots & \\ \mathbf{0} & & & & \lambda_{mn} \end{bmatrix} \quad \lambda_{ij} = \begin{bmatrix} \cos\theta_{zZ} & 0 & 0 \\ 0 & \cos\theta_{xX} & \cos\theta_{xY} \\ 0 & \cos\theta_{yX} & \cos\theta_{yY} \end{bmatrix}$$

where $\theta$ is the intersection angle between the local and the global coordinate axes.

The structural global stiffness, mass matrix $\mathbf{K}_{mn}$, the global loading column vectors $\mathbf{f}_{mn}$, $\mathbf{F}_{mn}$ can be obtained by splicing

## 6. Numerical example

**Example 1.** A simply supported square plate with the geometric configuration of length $L$, the thickness $h$, the Poisson's ratio $\mu$ and subjected to the uniform loading $q$. Evaluate the deflections at the center point of the plate.

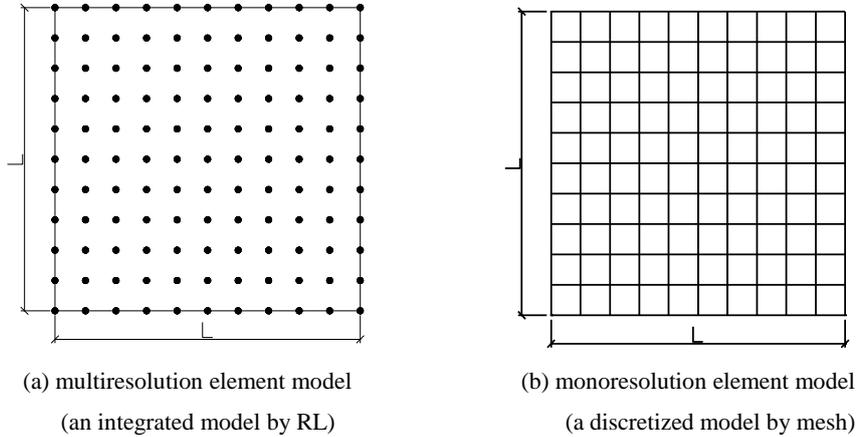

(a) multiresolution element model  (b) monoresolution element model
(an integrated model by RL)   (a discretized model by mesh)

Fig.6. Finite element models for the square plate

The displacement responses are found by the proposed element model (multiresolution element model), the corresponding traditional 4-node locking-free Mindlin plate element model (monoresolution element model) displayed in Fig.6. a,b, and the wavelet element based on two-dimensional tensor product B-spline wavelet on the interval (BSWI) [2] respectively. The BSWI is chosen because it is the best one among all existing wavelets in approximation of numerical calculation [15] and directly constructed by the tensor product of the wavelets expansions at each coordinate. The central deflections of the plate with the different thickness length ratios under the boundary conditions of four-side simply supported and four-side clamped are summarized in the table.1. One proposed element with the RL of $11\times 11$ is adopted and the 4-node conventional Mindlin rectangular plate element under the corresponding meshes of $10\times 10$ is also employed

**Table.1**. the deflection ($w/ qL^4 /100C_b$) at the center point of the simply- supported plate

| h/L | Four-side simply supported | | | | Four-side clamped | | | |
| --- | --- | --- | --- | --- | --- | --- | --- | --- |
| | analytic | paper | BSWI23 | BSWI43 | analytic | paper | BSWI23 | BSWI43 |
| $10^{-3}$ | 0.4062 | 0.4105 | 1.84E-4 | 0.4010 | 0.1265 | 0.1290 | 3.8E-5 | 0.1115 |
| $10^{-2}$ | 0.4062 | 0.4106 | 0.0173 | 0.4067 | 0.1265 | 0.1293 | 0.0037 | 0.1269 |
| $10^{-1}$ | 0.4273 | 0.4304 | 0.3510 | 0.4273 | 0.1499 | 0.1522 | 0.1152 | 0.1505 |
| 0.15 | 0.4536 | 0.4566 | 0.4152 | 0.4536 | 0.1798 | 0.1802 | 0.1592 | 0.1788 |
| 0.20 | 0.4906 | 0.4933 | 0.4678 | 0.4904 | 0.2167 | 0.2183 | 0.2047 | 0.2172 |
| 0.30 | 0.5956 | 0.5968 | 0.5861 | 0.5957 | 0.3227 | 0.3235 | 0.3187 | 0.3246 |
| 0.35 | 0.6641 | 0.6631 | 0.6579 | 0.6641 | 0.3951 | 0.3902 | 0.3898 | 0.3937 |

In the table.1. a 2D BSWI element of the jth scale=3, the mth order =2, 4 is used respectively abbreviated as BSWI23, BSWI43 with the DOF of $9\times 9$ and $11\times 11$. It can be seen that the analysis accuracies with the proposed element and the conventional element are the same and 2D BSWI element are gradually improved with the order reaching high. and the RL adjusting is quite easier than the order changing. Although the BSWI43 is of high accuracy, when compared with the proposed, the deficiencies of the BSWI element are obvious as follows. In light of tensor product formulation of the multi-dimensional MRA framework, the DOF of a multi-dimensional BSWI element will be so drastically increased from that of a one-dimensional element in an irrational way, resulting in substantial reduction of the computational efficiency. Secondly, there exists no such a parameter as the RL with a clear mathematical sense to adjust the element node number. As to the traditional monoresolution and the proposed multiresolution, the RL adjusting is more rationally and efficiently to be implemented than the meshing and the re-meshing to modulate element node number for the following two reasons. Firstly, the RL is based on the MRA framework which is constructed on a rigorous mathematical basis and the full node shape functions result in an integrated model, while the mesh, which resorts to the empiricism and the element split node shape functions lead to a discretized model, has no MRA framework. Secondly, the stiffness matrix and the loading column vectors of the proposed element can be obtained automatically around the nodes while those of the traditional 4-node rectangular plate elements acquired by the artificially complex reassembling around the elements. It can be seen that the computational accuracy and efficiency of the proposed element model is higher than the other two.

**Example 2.** Two square plates are subjected to the uniform loading *q* with the geometric configuration of length *L*, thickness *h*, Poisson's ratio $\mu$ and the ratios of L/h=$10^{-3}$ and L/h=0.3 respectively. Evaluate the deflections and moments at the center point of two plates under such boundary conditions as four side simply supported or clamped.

The displacement responses and moments are evaluated by the proposed multiresolution element method, the traditional monoresolution element method. The central deflections and the bending moments for a thin and a thick plates with the boundary conditions as simply supported four edges (SS) and the clamped four edges (SC) under the different RLs and meshes are displayed respectively in the table.2 and the table.3.The RL of the proposed and the corresponding meshes of the conventional are compared. It can be seen that the analysis accuracies with the proposed and the conventional are gradually improved respectively with the RL reaching high and the mesh approaching dense. However, the RL adjusting is more rationally and efficiently to be implemented than the meshing and remeshing to modulate element node number because the RL adjusting is based on the MRA framework which is constructed on a rigorous mathematical basis while the meshing or remeshing, which resorts to the empiricism, has no MRA framework. Thus, the computational efficiency of the proposed element method is higher than the traditional one. In this way, the proposed element exhibits its strong capability of accuracy adjustment and its high power of resolution to identify details (nodes) of deformed structure by means of modulating its resolution level, just as a multi-resolution camera with a pixel in its taken photo as a node in the proposed element. There appears no mesh in the proposed element just as no grid in the image. Hence, an element of superior analysis accuracy surely has more nodes when compared with that of the inferior just as a clearer photo contains more pixels.

**Table.2**. the deflection for thin plate (h/L=10$^{-3}$) under different RLs and meshes

| RL | Mesh | Central deflection (q$l^4$/100 $C_b$) | | | | Bending moment (q$l^2$/10) | | | |
|---|---|---|---|---|---|---|---|---|---|
| | | SS | | SC | | SS (central) | | SC(side middle) | |
| | | Multi | Mono | Multi | Mono | Multi | Mono | Multi | Mono |
| 3×3 | 2×2 | 0.5063 | 0.5063 | 0.1480 | 0.1480 | 0.6602 | 0.6602 | -0.3551 | -0.3551 |
| 5×5 | 4×4 | 0.4328 | 0.4328 | 0.1403 | 0.1403 | 0.5217 | 0.5217 | -0.4761 | -0.4761 |
| 9×9 | 8×8 | 0.4129 | 0.4129 | 0.1304 | 0.1304 | 0.4892 | 0.4892 | -0.5028 | -0.5028 |
| 17×17 | 16×16 | 0.4079 | 0.4079 | 0.1275 | 0.1275 | 0.4814 | 0.4814 | -0.5104 | -0.5104 |
| Analytical[14] | | 0.4062 | | 0.1265 | | 0.4789 | | -0.5133 | |

**Table.3**. the deflection for thick plate (h/L=0.3) under different RLs and meshes

| RL | Mesh | Central deflection (q$l^4$/100 $C_b$) | | | | Bending moment (q$l^2$/10) | | | |
|---|---|---|---|---|---|---|---|---|---|
| | | SS | | SC | | SS (central) | | SC(side middle) | |
| | | Multi | Mono | Multi | Mono | Multi | Mono | Multi | Mono |
| 3×3 | 2×2 | 0.7081 | 0.7081 | 0.3806 | 0.3806 | 0.8395 | 0.8395 | -0.4089 | -0.4089 |
| 5×5 | 4×4 | 0.6201 | 0.6201 | 0.3377 | 0.3377 | 0.5739 | 0.5739 | -0.4604 | -0.4604 |
| 9×9 | 8×8 | 0.5992 | 0.5992 | 0.3250 | 0.3250 | 0.5031 | 0.5031 | -0.4604 | -0.4604 |
| 17×17 | 16×16 | 0.5942 | 0.5942 | 0.3219 | 0.3219 | 0.4850 | 0.4850 | -0.4546 | -0.4546 |
| Analytical[14] | | 0.5956 | | 0.3227 | | 0.4789 | | -0.4260 | |

## 7. Discussion

From the numerical example above, it is shown that based on the multi-resolution rectangular Mindlin plate element formulation, the multi-resolution finite element method is introduced, which incorporates such main steps as RL adjusting, element matrix formation, element matrix transformation from a local coordinate system to a global one and global structural matrix formation by splicing of the element matrices. Owing to the existence of the new MRA framework, the RL adjusting for the proposed method is more rationally and easily to be implemented than the meshing and re-meshing for the traditional 4-node rectangular Mindlin plate element method. Due to the basic node shape function, the stiffness matrix and the loading column vectors of a proposed element can be automatically acquired through quadraturing around nodes in the element matrix formation step while those of the traditional 4-node rectangular Mindlin plate element obtained through complex artificially reassembling of the element matrix around the node-related elements in the re-meshing process for their 1/4 split nodes in a conventional element, which contributes a lot to computation efficiency improvement of the proposed method. Moreover, since the multiresolution rectangular Mindlin plate element model of a structure usually contains much less elements than the traditional element model, thus requiring much less times of transformation matrix multiplying, the computation efficiency of the proposed method appears much higher than the traditional in the step of element matrix transformation. In addition, because of the simplicity and clarity of a full node shape function formulation with the Kronecker delta property and the solid mathematical basis for the new MRA framework, the proposed method is also superior to other corresponding MRA methods in terms of the computational efficiency, the application flexibility and extent. Hence, taking all those causes into account, the conclusion can be drawn that the multi-resolution Mindlin rectangular plate element method is more rationally, easily and efficiently to be executed, when compared with the traditional 4-node Mindlin rectangular plate element method or other corresponding MRA methods, and the proposed element would be the most accurate one formulated ever since.

In addition, multiresolution analysis (MRA) can be viewed as a technique by which amount of element details that are exposed can be modulated at a request. In the numerical analysis field, the node number a large-sized element contains could be adjusted respectively in various manners by different methods, such as the traditional FEM, the wavelet FEM (WFEM), the traditional meshfree method (MFM), the traditional natural element method (NEM), and the proposed multiresolution FEM (MFEM) etc. FEM applies the scheme of meshing and re-meshing, which is mainly relied on the empiricism, to adjust the element node number in a rough way, thus performing an irrational MRA; WFEM adopts the technique based on a mutually nested subspace sequence that is a weak mathematical basis and quite complex to be founded. MFM and NEM employ the strategy of prior artificial-selected element node layout which is also largely dependent on the empiricism. In a word, all those above or other methods are short of the parameter-resolution level (RL) with a clear mathematical sense that can be easily used to fully alter total element node number and locate element node because they do not have a simple, clear and solid mathematical basis. However, MFEM has such a simple, clear and rigorous mathematical basis that brings about the parameter RL to freely adjust total node number and locate nodes within the element. Hence, it can be said that WFEM, MFM, NEM etc are the intermediate products in the transition of the traditional FEM from the monoresolution to the multiresolution and MFEM consolidates all these irrational MRA approaches.

## 8. Conclusions and prospective

A new multiresolution finite element method that has both high power of resolution and strong flexibility of analysis accuracy is introduced into the field of numerical analysis. The method possesses such prominent features as follows:

i. A novel technique is proposed to construct the simple and clear basic node shape function for a locking-free Mindlin rectangular plate element, which unveils the secret behind assembling artificially of node-related items in global matrix formation by the conventional FEM.

ii. A mathematical basis for the MRA framework, that is the mutually nesting displacement subspace sequence, is constituted out of the scaled and shifted version of the basic node shape function, which brings about the rational MRA concept together with the RL.

iii. The traditional 4-node Mindlin rectangular plate element and method is a monoresolution one and

also a special case of the proposed. An element of superior analysis clarity surely contains more nodes when compared with that of the inferior.

iv. The RL adjusting for the multiresolution Mindlin plate element model is laid on the rigorous mathematical basis while the meshing or remeshing for the monoresolution is based on the empiricism. Hence, the implementation of the proposed element method is more rational and efficient than that of the traditional or other MRA methods such as the wavelet finite element method, the traditional meshless method, and the traditional natural element method etc. The proposed element method can consolidate all these corresponding irrational MRA approaches. Thus, the accuracy of a plate structural analysis is replaced by the clarity, the irrational MRA by the rational the mesh by the RL that is the discretized model by the integrated.

v. A quite new concept is introduced into the FEM that the structural analysis clarity is actually determined by the RL-the density of node uniform distribution, not by the mesh.

vi. With advent of the new finite element method [16, 17, 18], the rational MRA will find a wide application in numerical solution of engineering problems in a real sense.

The upcoming work will be focused on the treatment of interface between multiresolution elements of different RL. The interface may be extended to the bridging domain in which a transitional element could be used just as PS images of different RL. The transitional element could also be constructed by the technique of scaling and shifting of the basic full node shape function to virtual or real nodes.